\documentclass[twoside,twocolumn,showpacs,prb]{revtex4}
\usepackage{graphicx}
\usepackage{amsmath}
\usepackage{amssymb}

\begin{document}
\title{Critical Currents and Order-Disorder Phase Transition in the Vortex States of YBa$_2$Cu$_4$O$_8$ with Chemically Introduced Disorder}
\author{M. Angst}
 \email[Corresponding author. Email: ]{angst@phys.ethz.ch}
\author{S. M. Kazakov}
\author{J. Karpinski}
\affiliation{Solid State Physics Laboratory ETH, CH 8093-Z\"urich,
Switzerland}
\author{A. Wisniewski}
\author{R. Puzniak}
\author{M. Baran}
\affiliation{Institute of Physics, Polish Academy of Sciences, Al.
Lotnikow 32/46, PL 02-668 Warsaw, Poland}
\date{\today}
\begin{abstract}
A series of YBa$_{2-x}$Sr$_x$Cu$_4$O$_8$ single crystals was
measured to study the influence of site disorder on the transition
line $H_{\text{ss}}(T)$ between quasi-ordered vortex lattice and
highly disordered vortex glass, as well as on the maximum critical
current density within the glass phase,
$j_{\text{c}}^{\text{max}}$. When 32\% of Ba is replaced by Sr,
$j_{\text{c}}^{\text{max}}$ is an order of magnitude higher than
in the unsubstituted compound. In contrast, the transition field
$H_{\text{ss}}$ first drops by a factor of about five with a
substitution of just 10\% of Sr for Ba, and then remains
approximately constant for higher Sr contents. Our results
indicate that in very clean systems the order-disorder transition
is affected very strongly by any crystallographic disorder, while
above a certain threshold it is relatively robust with respect to
additional disorder. In all substituted crystals $H_{\text{ss}}$
monotonically decreases with an increase of temperature.
\end{abstract}
\maketitle

\section{Introduction}
\label{intro} The vortex matter of cuprate superconductors has a
rich phase diagram in the $H$-$T$ plane (see e.g.\ Ref.\
\onlinecite{Blatter94}). Not long after the discovery of the
cuprate superconductors it has been established that there are at
least two phases, a fluid (liquid or gas)\cite{Sasagawa98} phase
at high temperatures and fields and a solid (lattice or glass)
phase at low temperatures and fields. It was shown that the
transition between these two phases is of first order, at least in
low fields up to a critical point.
\cite{Safar92,Cubitt93,Schilling96} Recently, it has been
shown\cite{Khaykovich96,Khaykovich97,Deligiannis97,Giller97,Nishizaki98,Giller99,Nishizaki00,Radzyner00}
that there are two distinct solid phases. In low fields the
vortices form a quasi-ordered lattice in a so-called Bragg
glass,\cite{Nattermann90,Giamarchi94} which is stable against the
formation of dislocations. In higher fields they form a highly
disordered, entangled\cite{Blatter94} solid. The nature of the
high field phase is not well understood. It could be anything from
a highly viscous fluid to a vortex glass with unbounded barriers
against vortex movement.\cite{Fisher89,Reichhardt00} We will
simply term it ``glass'' henceforth.

The high-field glass phase is caused by weak, random and
uncorrelated disorder due to point-like defects, such as oxygen
vacancies.\cite{Blatter94} This type of disorder is present even
in the most pure crystals of cuprate superconductors. Strong,
correlated disorder due to extended defects, such as twin
boundaries or columnar irradiation defects, has got a different
influence on the vortex matter and can lead to additional phases,
e.g.\ a ``Bose glass" phase\cite{Nelson92} located at low
fields.\cite{Kwok00} Correlated disorder may not be present in all
crystals. In the following, when we simply speak of ``disorder'',
we are referring to weak random point-like disorder.

The pinning induced by the disorder has two effects on the vortex
lattice.\cite{Vinokur98} Firstly, it localizes vortices by
trapping them in low-lying metastable states, hardening the solid.
Secondly, it promotes transverse wandering of vortices from their
ideal lattice positions. The mechanism of the destruction of the
lattice, triggering the transition to the glass
phase,\cite{Giamarchi94,Vinokur98,Ertas96,Giamarchi97,Koshelev98}
is very similar to the mechanism of melting at higher temperature.
Similar to the way the lattice looses translational order and
melts into a fluid, when thermal fluctuations become of the order
of the lattice spacing, it looses translational order and
``melts'' into a glass, when the disorder-induced line wandering
becomes of the order of the lattice spacing. In both cases the
stability against the formation of topological defects, such as
dislocations, is lost. The massive proliferation of dislocations
accompanying the destruction of translational order leads to an
entangled configuration of vortices, i.e.\ topological order is
lost as well. In highly anisotropic superconductors, the
lattice-glass transition may fall together with a decoupling
between the layers,\cite{Giamarchi97,Koshelev98} of which there is
some evidence in Bi$_2$Sr$_2$CaCu$_2$O$_{8+\delta}$
(BSCCO).\cite{Gaifullin00} This is not expected in less
anisotropic systems, like YBa$_2$Cu$_3$O$_{7-\delta}$
(Y123).\cite{Giamarchi97,Koshelev98} There is considerable
evidence that the phase transition between lattice and glass, like
the melting transition between lattice and fluid, is of first
order\cite{Radzyner00,Gaifullin00,Giller00,Paltiel00} and is even
a continuation of the melting line.\cite{Avraham01}

Since the solid-solid transition is an order to disorder
transition with the disordered phase located at high fields, the
transition line $H_{\text{ss}}(T)$ can be expected to shift to
lower fields with the introduction of additional disorder in the
crystal structure.
Calculations\cite{Vinokur98,Ertas96,Giamarchi97,Koshelev98} of
$H_{\text{ss}}$, based on a Lindemann criterion, confirm this
expectation. Indeed, there seems to be a correlation between the
position of $H_{\text{ss}}$ and the purity of a sample, as can be
seen by comparing different measurements on Y123 crystals. Also,
studies of the effect of electron irradiation on the transition
between lattice and glass have found a systematic decrease of
$H_{\text{ss}}$ with increasing irradiation
dose.\cite{Khaykovich97,Nishizaki00}

An alternative way to introduce structural disorder into a system
is to partially substitute an element of the compound. However,
the generally varying level of oxygen within samples, which is
difficult to measure and control, constitutes a serious problem.
YBa$_2$Cu$_4$O$_8$ (Y124) is a compound very well suited for
studying the dependence of $H_{\text{ss}}$ on chemically
introduced disorder, because in this compound the oxygen content
is fixed to 8 per unit cell. Another advantage of Y124 is that
twinning, which leads to strong correlated disorder, does not
exist in this compound.\cite{Karpinski99} Finally, Y124 has got an
intermediate anisotropy $\gamma \approx 12$,\cite{Karpinski99}
which, as compared to Y123, shifts the glass phase down to fields
more easily attainable experimentally. To exclude or at least to
minimize concurrent sources of disorder in the system it is
advisable to use single crystals for any measurements.

If vortices are not ordered in a lattice, they can better adapt to
the local pinning potential. Therefore, critical current densities
can be expected to be higher in the glass phase. High critical
currents in high magnetic fields are desirable for many
applications. Investigations of the disordered solid state and
it's boundaries are thus very important in view of practical
applications. Many irradiation studies found that critical current
densities can increase by orders of magnitude upon irradiating a
sample.\cite{Dover89} Studies on compounds with chemical disorder
have also found increases of $j_{\text{c}}$ and even the
development of a second peak.\cite{Murakami94}

In this work, we study the influence of a substitution of
isovalent Sr for Ba in YBa$_{2-x}$Sr$_x$Cu$_4$O$_8$ on the lattice
to glass transition line $H_{\text{ss}}(T)$, the irreversibility
line $H_{\text{irr}}(T)$ and the critical current densities
$j_{\text{c}}$. In the paper, we are going to give evidence of an
enhancement of the maximum critical current density in  the glass
phase $j_{\text{c}}^{\text{max}}$ by more than an order of
magnitude due to structural disorder introduced by Sr
substitution. The change of the vortex matter phase diagram due to
the influence of relatively strong quenched disorder will be
presented.

\section{Experimental}
\label{exp}

\label{samples} 
\begin{figure}
\includegraphics[width=0.95\linewidth]{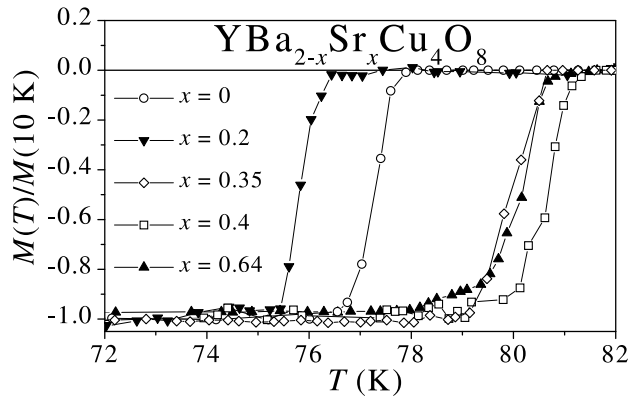}
\caption{Normalized $M(T)$ curves of the crystals used in this
work showing sharp superconducting transitions. The curves shown
were measured in a magnetic field of $H=1 \, {\text{Oe}}$, under
zero field cooled conditions.} \label{Fig1}
\end{figure}

Single crystals of Y124, with and without Sr
substitution, were grown using a high-pressure technique, the
details of crystal growth are reported
elsewhere.\cite{Karpinski00a} Single crystals of
YBa$_{2-x}$Sr$_x$Cu$_4$O$_8$ with Sr content up to $x=0.64$ or
32\% were obtained. The crystals were checked with EDX and their
structure analyzed with Single Crystal x-ray
analysis.\cite{Karpinski00a} The transition temperatures
$T_{\text{c}}$ of the crystals were determined with a SQUID
magnetometer by temperature sweeps (both zero field and field
cooled) in an applied field of $1 \, {\text{Oe}}$. All samples
chosen for further magnetic measurements had a well-resolved
transition with a transition width (10\% to 90\%) smaller than $2
\, {\text{K}}$ (see Fig.\ \ref{Fig1}).

\label{magn_exp} Measurements of the dc magnetization were
performed with a Quantum Design MPMS5 SQUID magnetometer,
additional measurements were performed with a non-commercial SQUID
magnetometer with a sensitivity better than $10^{-6} \,
{\text{emu}}$. All measurements were done with the magnetic field
applied parallel to the $c$-axis of the crystal and generally the
magnetization was measured at fixed temperatures as a function of
the external field being swept up and down. After each change of
the field, we waited for a time of $5 \, {\text{min}}$ before
measuring the magnetic moment four times using a scan length of $2
\, {\text{cm}}$. The relatively short scan length was chosen to
ensure maximum homogeneity of the applied field and was necessary
to ensure a sufficient resolution of all features of interest
discussed below. The wait time of $5 \, {\text{min}}$ was chosen
to avoid effects of the initial relaxation of the field of the
superconducting magnet without extending the time of measurements
too much. Throughout the paper $H$ denotes the applied field,
i.e.\ no correction of the demagnetization was made. This is not a
problem, however, since the magnetization is always much smaller
than the external field except in the low field range, which is
not the particular interest in this paper.

The critical current density $j_{\text{c}}$\cite{notejcjs} was
calculated from the width of the hysteresis loop using an extended
Bean model.\cite{Wiesinger92} The irreversibility field
$H_{\text{irr}}$ was determined as the field where increasing and
decreasing field branches of the $M(H)$ hysteresis loop meet, or
equivalently as the field where the critical current density
vanishes. A uniform criterion of $j_{\text{c}} = 50 \,
{\text{Acm}^{-2}}$ was used. For the crystals studied it
corresponds to a criterion of in between $4.4 \, {\mathrm{\mu}
\text{emu}}$ and $26 \, {\mu \text{emu}}$, which is not far from
the accuracy limit of our experimental setup.

\section{Results and Discussion}
\label{res}

In the following we will discuss the influence of Sr substitution
on disorder and structure, critical current densities, the
irreversibility line, and the transition from lattice to glass.

\subsection{Disorder and Structure}
\label{Sr_disc} Sr substitution increases the disorder
in the structure, evidenced for example by a dramatically
increased NQR line width.\cite{Karpinski00a} No evidence for an
introduction of correlated disorder was found and since twin
boundaries are absent in YBa$_{2-x}$Sr$_x$Cu$_4$O$_8$ the
assumption of the disorder in our crystals to be purely random and
point-like seems justified.

The Sr substitution also induces a variation of structural
parameters, causing a charge redistribution. Bond valence sum
calculations and nuclear quadrupole resonance measurements
indicate a transfer of holes from oxygen to copper atoms in the
CuO$_2$ planes and from copper to oxygen in the
chains.\cite{Karpinski00a} The charge redistribution can explain
the increase of $T_{\text{c}}$ upon substituting more than 10\% Sr
for Ba. The initial decrease of $T_{\text{c}}$ may be due to the
increased site disorder.\cite{Attfield98}

An important structural modification due to Sr substitution is a
decrease of the thickness of the blocking layer $d_{\text{b}}$
separating the superconducting CuO$_2$ planes, linear with Sr
content. A substitution of 32\% Sr for Ba causes the blocking
layer to shrink by about 1\%,\cite{Karpinski00a} which leads to an
improved coupling between the superconducting planes. Torque
measurements indeed indicate a corresponding decrease of the
anisotropy $\gamma$.\cite{ownfutur}

\subsection{Critical Current Densities}
\label{jc_disc} 
\begin{figure}
\includegraphics[width=0.95\linewidth]{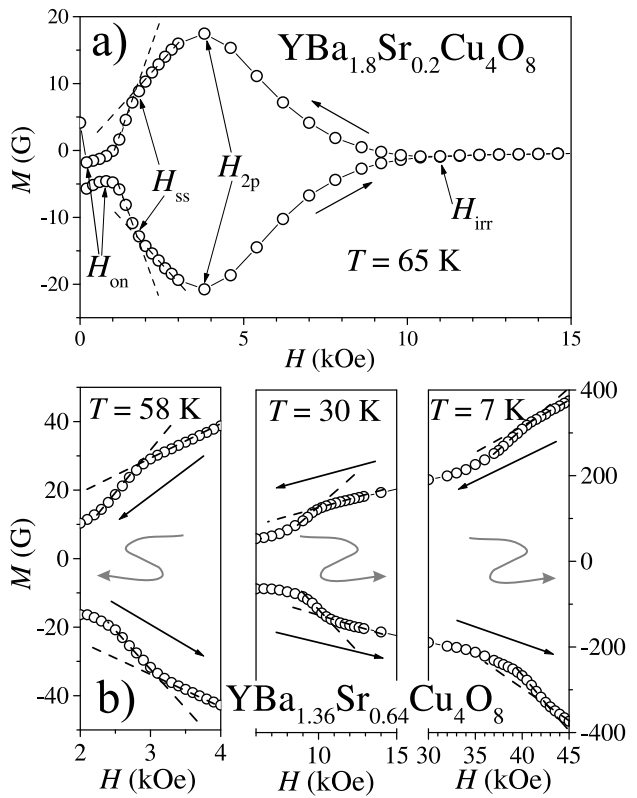}
\caption{a) One of the hysteresis loops measured. The direction of
the field change is shown by arrows. The onset of the second peak,
the second peak maximum and the kink in $M(H)$ between them are,
denoted as $H_{\text{on}}$, $H_{\text{2p}}$ and $H_{\text{ss}}$,
respectively. Note that there are small differences between the
respective fields between the two directions of field change (see
text and Fig.\ \ref{Fig10}). The irreversibility field, where
field increasing and decreasing branches meet, is denoted as
$H_{\text{irr}}$. b) Additional parts of hysteresis loops showing
the kink region at three different temperatures.} \label{Fig2}
\end{figure}

Figure \ref{Fig2}a) shows one of the half
hysteresis loops measured. The second peak is clearly discernible
and dominates $M(H)$. In all crystals, the second peak was
observed over the whole temperature range measured, even just $1
\, {\text{K}}$ below $T_{\text{c}}$. The form of the hysteresis
loop and thus of $j_{\text{c}}(H)$ can be linked to the different
phases of vortex matter. In low fields hysteresis is small and the
hysteresis loop is rather anisotropic, which indicates that the
major source of hysteresis in low fields is not of bulk origin
(see also Sec.\ \ref{IL_disc}). The origin of the low bulk
hysteresis is probably that the vortices are in the lattice phase
in small $H$. Weak random point disorder is not effective in
pinning an ordered dislocation-free lattice of vortices. The
magnitude of the critical current density in low fields gives an
indication of the degree of disorder in the system. For our
samples, measurements of the critical current density at reduced
temperatures of $T/T_{\text{c}}=0.4$ and $0.7$ in fields of $1$
and $2\, {\text{kOe}}$ suggest that the disorder is increased
monotonically up to the highest substitution level measured, with
the biggest rate between $x=0.4$ and $x=0.64$.

When, upon rising the field, the lattice is destroyed in the glass
state and the vortex system has an additional ``dislocation degree
of freedom'',\cite{Giamarchi97} individual vortices can much
better adapt to local minima in the pinning potential. This
results in a higher effective pinning force and thus a higher
critical current density $j_{\text{c}}$ and relaxation
barrier.\cite{Koshelev98} Additionally, the entanglement itself
was also suggested to increase critical current densities, due to
an increased intervortex viscosity,\cite{Marchetti91} provided the
barriers to flux cutting are high
enough.\cite{Ertas96,Giamarchi97} In our case, a steep rise of the
magnetization occurs just below the kink field, denoted
$H_{\text{ss}}$ in Fig.\ \ref{Fig2}a) (see also Sec.\
\ref{Hss_disc}). Local magnetic measurements on a BSCCO crystal
revealed a sharp increase in $|M(H)|$, followed by a monotonic
decrease afterwards.\cite{Khaykovich96} The so-called ``second
peak'' (or ``fishtail'') effect in BSCCO can thus be accounted for
by the order-disorder transition. On untwinned Y123 crystals, a
similar sharp increase in $|M(H)|$ was observed by local magnetic
measurements,\cite{Giller99} but contrary to the situation for
BSCCO, $|M(H)|$ continues to increase upon increasing field,
although with a slower rate, which leads to a sharp kink in
$M(H)$. Only in even higher fields the magnetization starts to
decrease with increasing field.

Independently of vortex matter phase transitions, the second peak
or fishtail effect in Y123 was attributed to a crossover from
elastic to dislocation-mediated plastic
creep.\cite{Deligiannis97,Abulafia96} This mechanism is closely
connected to the lattice-glass transition though, since in the
lattice phase dislocations are strongly suppressed while in the
glass phase dislocations proliferate, as mentioned
above.\cite{note2pbelowHm} From collective creep theory a rise of
the experimentally observable critical current density with the
field is expected,\cite{Blatter94,Abulafia96} as long as elastic
creep can be assumed and the bulk pinning is relevant. From the
shape of the hysteresis loops shown in Fig.\ \ref{Fig2} follows
that in YBa$_{2-x}$Sr$_x$Cu$_4$O$_8$ the situation is similar to
the one in  Y123. After rising the field even more, above
$H_{\text{irr}}$, the vortex matter finally enters the fluid
phase, where critical current densities are zero.

The dependence of the hysteresis loop of an unsubstituted Y124
single crystal on defects created by fast neutron irradiation was
studied in Ref.\ \onlinecite{Werner00}. Fast neutron irradiation
leads to improved pinning properties due to collision cascades
(strongly pinning extended defects of spherical shape) and
agglomerates of smaller defects. A clearly discernible second peak
was observed in the untreated crystal. In the weakly irradiated
crystal the second peak was less pronounced and after irradiation
to a fluence of $10^{17}\, {\text{cm}}^{-2}$ it disappeared. At
the same time, $j_{\text{c}}$ drastically increased in low to
medium fields and the hysteresis loops became more symmetrical.

The effect of fast neutron irradiation to remove the second peak
may be understood from the different defect structure. The
disorder resulting from the collision cascades is to some extent
correlated and fast neutron irradiation can therefore create a
Bose glass phase in low fields. Bulk pinning is also strong in the
Bose glass phase, and if the irradiation induced correlated
disorder is strong enough, the Bose glass phase may well destroy
all of the Bragg glass phase. In high fields fast neutron
irradiation is not as effective in enhancing $j_{\text{c}}$ and
the irreversibility fields $H_{\text{irr}}$ are not
increased.

\begin{figure}
\includegraphics[width=0.95\linewidth]{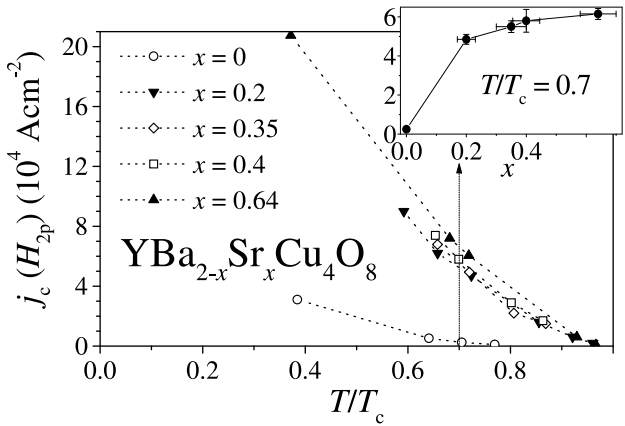}
\caption{Critical current density at $H_{\text{2p}}$ vs.\ reduced
temperature, for different Sr contents. It can be seen that
unsubstituted Y124 has much lower maximum critical current
densities than substituted Y124, at all temperatures. Inset:
$j_{\text{c}}(H_{\text{2p}})$ vs.\ Sr content $x$ at
$T/T_{\text{c}}=0.7$.} \label{Fig3}
\end{figure}

\begin{figure}
\includegraphics[width=0.95\linewidth]{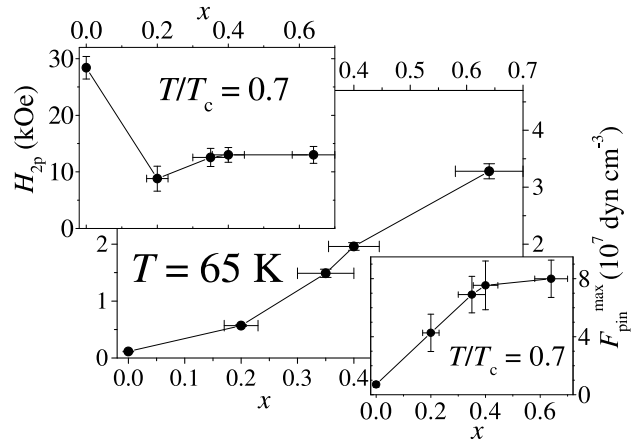}
\caption{Dependence of the maximum pinning force density
$F_{\text{pin}}^{\text{max}}=H_{\text{2p}}
j_{\text{c}}(H_{\text{2p}})$ on Sr content, at a temperature of
$65 \, {\text{K}}$. The maximum pinning force density increases by
more than an order of magnitude upon substituting 32\% of Sr for
Ba. Upper left inset: Second peak field vs.\ Sr content at
$T/T_{\text{c}}=0.7$. Lower right inset: Maximum pinning force
density vs.\ Sr content, at $T/T_{\text{c}}=0.7$.} \label{Fig4}
\end{figure}

In our case of Sr substitution the significant fields shift, but
the qualitative form of the hysteresis loop remains the same. As
with fast neutron irradiation, Sr substitution increases
$j_{\text{c}}$, but most pronounced in medium fields, in the
second peak region. The maximum critical current densities there,
$j_{\text{c}}^{\text{max}}(T) \equiv
j_{\text{c}}(H=H_{\text{2p}},T)$, increase dramatically upon Sr
substitution, as can be seen in Fig.\ \ref{Fig3}. At the
temperature of $60 \, {\text{K}}$, the critical current density
reaches almost the values of critical current densities for Y123:
For the crystal with 32\% Sr substitution
$j_{\text{c}}^{\text{max}}(T=60 \, {\text{K}})=6 \times 10^4 \,
{\text{Acm}^{-2}}$, while for a Y123 single crystal
$j_{\text{c}}^{\text{max}}(T=60 \, {\text{K}})=6.3 \times 10^4 \,
{\text{Acm}^{-2}}$.\cite{Werner98} This is not the case, however,
at the lower temperature of $50 \, {\text{K}}$, where Y123
crystals have a maximum critical current density roughly three
times higher\cite{Werner00} than the one measured on Sr
substituted YBa$_{2-x}$Sr$_x$Cu$_4$O$_8$. The main panel of Fig.\
\ref{Fig4} shows the dependence on Sr content of the maximum
pinning force density $F_{\text{pin}}^{\text{max}}=H_{\text{2p}}
j_{\text{c}}(H_{\text{2p}})$ corresponding to the critical current
densities measured at $T=65 \, {\text{K}}$. It can be seen that
$F_{\text{pin}}^{\text{max}}$ increases by more than an order of
magnitude upon Sr substitution. The lower right inset of Fig.\
\ref{Fig4} shows the Sr dependence of
$F_{\text{pin}}^{\text{max}}$ at a fixed {\em reduced}
temperature. The initially approximately linear rise of the
maximum pinning force density saturates for $x \approx 0.4$
indicating that increasing the substitution level to $x = 0.64$
does not increase the pinning drastically any more. The critical
current density itself also shows saturation behaviour, but for
lower Sr content (see inset of Fig.\ \ref{Fig3}). The difference
is due to the substitution dependence of the second peak field,
shown in the upper left inset of Fig.\ \ref{Fig4}. The dependence
$H_{\text{2p}}(x)$ is similar to the substitution dependence of
$H_{\text{ss}}$, discussed in Sec.\ \ref{Hss_disc}

\subsection{Irreversibility Fields}

\label{IL_disc} 
\begin{figure}
\includegraphics[width=0.95\linewidth]{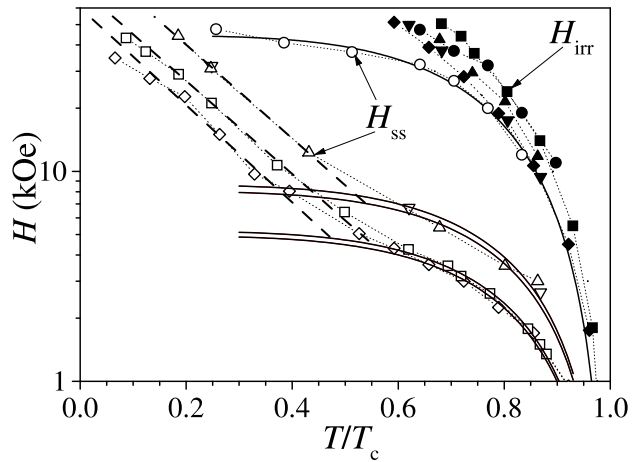}
\caption{The graph summarizes the dependence on the reduced
temperature of the irreversibility fields (full symbols) and
$H_{\text{ss}}$ (open symbols), for different Sr contents in
YBa$_{2-x}$Sr$_x$Cu$_4$O$_8$: circles correspond to $x=0$,
diamonds to $x=0.2$, down triangles to $x=0.35$, up triangles to
$x=0.4$ and squares to $x=0.64$. Error bars have been omitted for
clarity. Full lines are fits of the $H_{\text{ss}}(T)$ data to
Eq.$\:$(\ref{deltaTc}). Except for $x=0$, fitting was restricted
to the range $0.65 \leq T/T_{\text{c}} \leq 0.9$. Dashed lines are
guides for the eye indicating an exponential temperature
dependence of $H_{\text{ss}}$ at lower temperatures. Variations of
$H_{\text{irr}}(T)$ generally are relatively small and the details
depend on the temperature. $H_{\text{ss}}(T)$ of the unsubstituted
crystal is located in much higher fields than $H_{\text{ss}}(T)$
of all substituted crystals and for high temperatures is
practically located in the region of the irreversibility fields.}
\label{Fig5}
\end{figure}

The irreversibility fields $H_{\text{irr}}$ as a
function of temperature are, for different Sr substitution levels,
presented in Fig.\ \ref{Fig5}. The theoretical meaning of the
irreversibility field is not very clear, often parts of it are
controlled by surface or geometrical barriers (see below). In high
fields, it was often found to be located near the transition from
glass to fluid (see e.g.\ Refs.\
\onlinecite{Nishizaki98,Nishizaki00}). However, whether a true
phase transition between a glass and a fluid phase even exists, is
not a settled question. Not much is known about the properties of
the highly disordered entangled solid located above
$H_{\text{ss}}$, apart from properties that distinguish it from
the lattice phase. The feature differentiating the phase from the
liquid phase is a rapid freezing of the dynamics, but it is not
sure whether this freezing of the dynamics is a proper phase
transition or merely a crossover. The existence of a separate
so-called vortex glass (VG) phase, distinguished by an unbounded
distribution of the heights of barriers between metastable states,
was proposed early after the discovery of high $T_{\text{c}}$
superconductivity.\cite{Fisher89} Vortex glass scaling of current
voltage relations was used to determine the vortex glass
transition line $H_{\text{g}}$ experimentally, and also provided
strong support for the VG theory (see e.g.\ Refs.\
\onlinecite{Safar92,Nishizaki98}). However, later experiments and
numerical calculations found discrepancies (see Ref.\
\onlinecite{Reichhardt00} and references therein), and suggest a
glass phase more akin to window glass. Despite these
uncertainties, $H_{\text{irr}}(T)$ is certainly important from a
more practical point of view, as it limits the field range for
applications.

\begin{figure}
\includegraphics[width=0.95\linewidth]{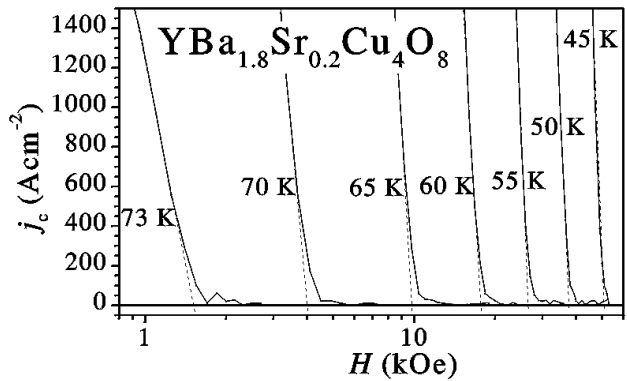}
\caption{Critical current density vs.\ logarithm of the magnetic
field, for the sample with 10\% Sr substitution. The relationship
is linear for high critical currents. Below a temperature
dependent threshold, deviations appear. Dashed lines extrapolate
the linear relationship to $j_{\text{c}}=0$.} \label{Fig6}
\end{figure}

It is hard to see a clear trend in the Sr substitution dependence
of the measured irreversibility fields. Differences between
$H_{\text{irr}}$ of the different samples measured are relatively
small, and the details depend on the temperature as well. Now it
is important to remember that bulk pinning is not the only source
of magnetic hysteresis. There are additional contributions due to
Bean-Livingston surface
barriers\cite{Bean_Livingston64,Burlachkov94} and due to
geometrical barriers.\cite{Zeldov94} The significance of surface
and geometrical barriers depends on several factors. One of them
is the strength of the bulk pinning - the smaller the (bulk)
critical current density the bigger the relative importance of
barriers. In YBa$_{2-x}$Sr$_x$Cu$_4$O$_8$ the critical current
densities depend, as discussed above, strongly on $x$, and are
very low for the unsubstituted crystal. Indeed, while the
hysteresis loops of substituted crystals are generally rather
symmetric, this is not the case for the unsubstituted crystal,
which is an indication that barrier hysteresis is more important
than bulk hysteresis.\cite{Burlachkov94} Another factor is the
sample shape - bulk hysteresis depends linearly on the sample
diameter, while the barrier hysteresis scales with the ratio of
thickness to diameter. Thus, barrier effects are more important
for thicker and smaller (in diameter) crystals. The thickness to
diameter ratio varies from $0.1$ to $0.3$ for the crystals used in
this study. Finally, an important factor is the temperature, since
the bulk hysteresis is depressed by thermal depinning, while the
one due to geometrical barriers is less affected by
temperature.\cite{Doyle98} In the second peak region, the main
part of the hysteresis is clearly of bulk origin, as follows from
the shape of the hysteresis loops (see Fig.\ \ref{Fig2}). However,
since the total hysteresis observed is simply the sum of bulk and
barrier hysteresis, the irreversibility field may still be
determined by surface or geometrical barriers, since the bulk
hysteresis goes to $0$ faster at higher
fields.\cite{Burlachkov94,Zeldov95b_Dewhurst96} In high fields,
bulk critical current densities were found to go to zero
exponentially.\cite{Yamauchi98} A linear relationship between
$j_{\text{c}}$ and $\ln(H)$ is indeed followed in the crystals we
measured. Figure \ref{Fig6} shows this relationship for
YBa$_{1.8}$Sr$_{0.2}$Cu$_4$O$_8$. The figure shows also, however,
that there are systematic deviations below a certain temperature
dependent threshold of critical current densities. For higher
temperatures the deviation appears at larger critical current
densities. This may indicate that the irreversibility lines are
indeed influenced considerably by barrier hysteresis at higher
temperatures.

On the other hand, measurements of the influence of fast neutron
irradiation on the high temperature part of the irreversibility
field on a Y124 single crystal found a slight decrease of
$H_{\text{irr}}$ upon irradiating the crystal.\cite{Werner97}
Irradiation may reduce Bean-Livingston type surface barriers, but
those are, like bulk pinning, strongly temperature dependent and
were generally not found to be important near $T_{\text{c}}$.
Geometrical barriers, on the other hand, should not change upon
irradiation. As the irradiation did change the irreversibility
field at high temperatures, it would indicate that
$H_{\text{irr}}$ at high temperatures can not be due to
Bean-Livingston or geometrical barrier hysteresis.

\begin{figure}
\includegraphics[width=0.95\linewidth]{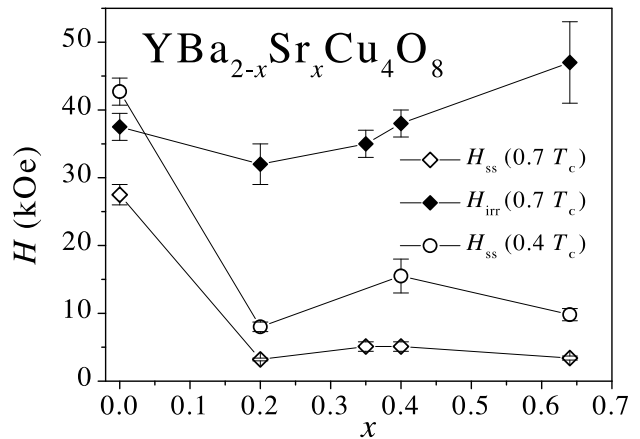}
\caption{Sr dependence of $H_{\text{ss}}$ at temperatures of $0.7
\, T_{\text{c}}$ (open diamonds) and $0.4 \, T_{\text{c}}$ (open
circles), and of $H_{\text{irr}}(0.7 \, T_{\text{c}})$ (full
diamonds). For both temperatures $H_{\text{ss}}$ drops
significantly by a small substitution of Sr for Ba, but remains
essentially constant upon further Sr substitution.
$H_{\text{irr}}$ also drops by a small Sr substitution, but less
significantly. It rises again upon further substitution and is
higher for $x=0.64$ than for $x=0$. The $x$ error bars have been
omitted for clarity.} \label{Fig7}
\end{figure}

However, as mentioned above, the importance of barriers strongly
depends on the sample shape and we cannot be completely sure to
what extent barriers influence $H_{\text{irr}}$, especially in the
case of the unsubstituted crystal, where bulk pinning is very
weak. In order to explain the effect of the Sr substitution on the
bulk properties, it is therefore advisable to compare
$H_{\text{irr}}(T)$ at not too high reduced temperatures. A
comparison of $H_{\text{irr}}(T)$ at $T=0.7 \, T_{\text{c}}$ is
shown in Fig.\ \ref{Fig7}. The position of the irreversibility
field is not changed much, but the general trend that
$H_{\text{irr}}$ rises with $x$ is consistent with the also rising
maximum critical current density. The only irreversibility field
that does not fit into this trend is the one of the unsubstituted
crystal, which may be attributed to barrier effects that still are
not negligible at this temperature for this weakly pinned
compound.

An increasing glass to liquid transition field can be justified
theoretically in a similar way as the decrease of the lattice to
glass transition field. Since the glass to liquid transition line
depends on the relation between the pinning energy and the thermal
energy, and because the pinning energy is increased upon
introducing more disorder into the system, it would seem natural
that the transition occurs at higher temperatures.

The dependence of the glass to fluid transition on disorder was
studied, for electron irradiated Y123, by Nishizaki {\em{et
al.}}.\cite{Nishizaki00} They measured both the glass transition
temperature $T_{\text{g}}$ resistively, and the irreversibility
field $H_{\text{irr}}$, and found the two lines $T_{\text{g}}(H)$
and $H_{\text{irr}}(T)$ to coincide. Nishizaki {\em{et al.}}\ find
the opposite disorder dependence of the glass to fluid transition,
in their case $H_{\text{g}}$ and $H_{\text{irr}}$ decrease with
increasing fluence of electron irradiation.

There are different possible explanations for the opposite
behaviour of the glass to liquid transition line when the disorder
is increased between the present measurements and those of
Nishizaki {\em{et al.}}.

Firstly, the disorder introduced by substituting 10\% or more Sr
for Ba in Y124 is much stronger than the disorder introduced by an
irradiation of $2.5 \, {\text{MeV}}$ electrons with a fluence of
up to $2 \times 10^{18} \, {\text{electrons/cm}^{2}}$ in untwinned
Y123. This can be seen by comparing the phase diagrams. The
irradiated Y123 crystals show a first order melting transition in
fields up to $50-100 \, {\text{kOe}}$, while all of our Sr
substituted Y124 crystals show no first order melting transition
at all up to temperatures very near $T_{\text{c}}$. Nishizaki
{\em{et al.}}\ note that the decrease of $H_{\text{g}}$ upon
increasing disorder they measured is consistent with a theoretical
prediction made for systems with weak pinning.\cite{Ikeda96} The
disorder in our substituted crystals may be too big for this to be
applicable.

Secondly, the rise of $H_{\text{irr}}$ with Sr substitution could
also be explained by a decreased anisotropy of the substituted
crystals, since $H_{\text{irr}}$ was found to scale with
$\gamma^{-2} s^{-1}$, where $s$ is the distance between two
adjacent superconducting planes.\cite{Sasagawa98} If the rise of
$H_{\text{irr}}$ in YBa$_{2-x}$Sr$_{x}$Cu$_4$O$_8$ with $x$ for
$x>0.2$ is due to a decreased anisotropy caused by the decreased
blocking layer thickness, the initial drop for $x<0.2$ may be due
to the disorder, in which case the dependence of $H_{\text{irr}}$
on disorder would be qualitatively the same as the one found by
Nishizaki \textit{et al.}.

\subsection{Transition from Lattice to Glass}

\label{Hss_disc} 
\begin{figure}[tb]
\includegraphics[width=0.95\linewidth]{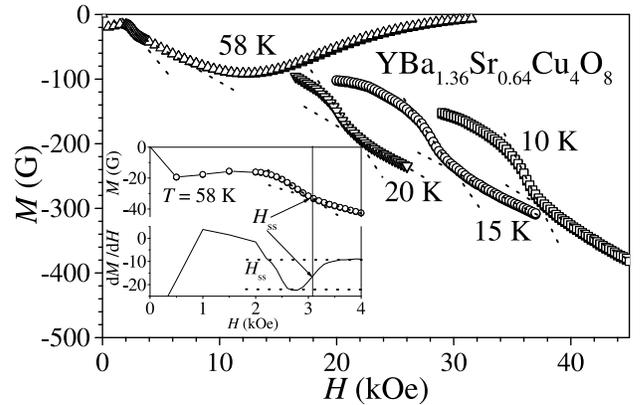}
\caption{$M(H)$ curves for increasing field near the onset of the
second peak. Dashed lines are guides for the eye pinpointing the
kink in the magnetization $H_{\text{ss}}$. The inset shows a
magnification of the onset region of a $M(H)$ curve and it's
derivative at $T=58\,{\text{K}}$. $H_{\text{ss}}$ is located at
the position of the steepest increase of $\text{d}M/\text{d}H$.}
\label{Fig8}
\end{figure}

As noted in Sec.\ \ref{jc_disc}, a kink in $M(H)$
(denoted $H_{\text{ss}}$) is visible between onset and maximum of
the second peak of the hysteresis loop shown in Fig.\ \ref{Fig2}.
According to previous studies performed on
Y123,\cite{Nishizaki98,Giller99,Nishizaki00} it is likely that the
kink corresponds to the transition line from the lattice to the
glass state of vortices. In Ref.\ \onlinecite{Radzyner00} it was
argued that the equilibrium phase transition corresponds to the
kink in $M(H)$ in decreasing field, while the kink in increasing
field corresponds to the upper limit of metastability of the
lattice phase. However, since the pinning is much stronger in the
glass phase, the equilibrium transition can be expected to be
located quite close to the upper limit of metastability. This was
indeed found from the measurements of Ref.\
\onlinecite{Radzyner00}. Also in our case, while we have more data
on the field increasing branch, the kink in $H$ decreasing was
checked at representative temperatures and the difference in the
two kinks was found to be rather small indeed (see Fig.\
\ref{Fig2}). We should mention that the actual equilibrium
transition is a little bit lower than the field of the kink in
$M(H^{\uparrow})$, which we call $H_{\text{ss}}$ (see Fig.\
\ref{Fig10}). It is interesting to note that generally the kink
positions (for both directions of field changes) do not depend on
the waiting time (we tested between $30\, \text{s}$ and $10\,
\text{min}$) after changing the the field. Additionally the kink
field location does not depend on the field change step size and
not on the small inhomogeneities of the magnetic fields ($\sim
0.02\%$) tested by using different scan lengths, but the {\em
shape} of the $M(H)$ curve around the kink {\em does} depend on
step size and scan length. On the other hand, the onset fields
$H_{\text{on}}$, defined as the fields where the absolute value of
the magnetization has a minimum, have a slight tendency to shift
to lower fields upon increasing waiting time, while the absolute
value of the magnetization at the onset is decreased significantly
upon increasing waiting time. Additionally the onset measured in
decreasing fields can be much lower (up to 50\% at low
temperatures) than the onset measured in increasing field. Figure
\ref{Fig8} shows a selection of magnetization curves measured, in
the region where the kink is located. The sharpness of the kink
varies with temperature and depends on the sample as well. The
inset of Fig.\ \ref{Fig8} shows the onset region of one of the
curves enlarged and the derivative of the magnetization (average
of the four scans), which can help to locate $H_{\text{ss}}$.
Within a range of a few Kelvin below $T_{\text{c}}$ the
unambiguous determination of $H_{\text{ss}}$ becomes difficult
nonetheless.

The increase of the absolute value of the magnetization around
$H_{\text{ss}}$ observed in our measurements is hardly sharp
enough to be called a ``jump''. One reason for this is the spatial
averaging of the inhomogeneous induction inside the sample.
Additionally, the difference between the onset of the second peak
in increasing fields $H_{\text{on}}^{\uparrow}$, defined as the
minimum of the absolute value of the magnetization before the
second peak, and the kink field $H_{\text{ss}}$ can be explained
by the existence of a region of metastability around the phase
transition.\cite{Radzyner00} Recent
experiments\cite{Giller00,Paltiel00} indicate that an abrupt
change of the field injects a transient disordered vortex phase at
the sample edges. If the thermodynamically stable phase is the
lattice phase, the transient disordered vortex phase then decays.
This decay happens with a rate decreasing to 0 as the field
reaches $H_{\text{ss}}$. In our measurements, the fields
$H_{\text{on}}^{\uparrow}$, $H_{\text{on}}^{\downarrow}$ and
$H_{\text{ss}}^{\downarrow}$ was found to follow the same
dependence on temperature and Sr content as $H_{\text{ss}}$
($H_{\text{ss}}^{\uparrow}$). Below the onset, both the magnitude
of the critical current density and the average value of the
magnetization for the two branches of the hysteresis loop suggest
that in this region surface and geometrical barriers are more
important than bulk pinning in our samples, i.e.\ bulk pinning
seems to be relevant only if the vortex matter is in the glass
phase.

The $H_{\text{ss}}(T)$ lines, as determined by the kink in the
$M(H^{\uparrow})$ curves for the measured samples, are also shown
in Fig.\ \ref{Fig5}. For all samples, $H_{\text{ss}}$ decreases
monotonically with increasing temperature. A monotonically
decreasing $H_{\text{ss}}$ was also found on
Nd$_{1.85}$Ce$_{0.15}$CuO$_{4-\delta}$,\cite{Giller97} but
measurements on
Y123\cite{Nishizaki98,Giller99,Nishizaki00,Radzyner00} showed an
$H_{\text{ss}}$ increasing with temperature. The position of the
phase transition between quasi-ordered lattice and highly
disordered glass is determined by the interplay between elastic
and pinning energy. For the case of weak, random disorder due to
point-like defects the theory of collective pinning was developed
by Larkin {\textit{et al.}}.\cite{Larkin72} The defects can
interact with the vortices in two ways.\cite{Blatter94} They can
cause a spatial variation of the transition temperature ($\delta
T_{\text{c}} $-pinning), described by a modulation of the linear
term of the Ginzburg-Landau free energy functional. Alternatively,
they can cause a spatial modulation of the mean free path ($\delta
\ell$-pinning), described by a modulation of the gradient term of
the free energy functional. In both cases, the influence of
disorder is described by a disorder parameter $\tilde{\gamma}$,
proportional to the defect density. However, the temperature
dependence of $\tilde{\gamma}$ is different for the two cases. For
$\delta T_{\text{c}} $-pinning, $\tilde{\gamma} \propto 1/
\lambda^4$, while for $\delta \ell$-pinning, $\tilde{\gamma}
\propto 1/ (\lambda \xi)^4$, where $\lambda$ and $\xi$ are the
penetration depth and the coherence length.\cite{Blatter94} The
order-disorder transition position was calculated analytically by
using a Lindemann criterion.\cite{Vinokur98,Ertas96} For the case
of not too big anisotropy and disorder, following the calculation
of Ref.\ \onlinecite{Vinokur98}, we get \begin{equation}
H_{\text{ss}} = H_{\circ} \left( \frac{U_{\circ}}{U_{\text{c}}}
\right) ^3 , \label{Hss_pos}
\end{equation} with $H_{\circ}=2 c_{\text{L}}^2 \Phi_{\circ} / \xi^2$,
$U_{\circ}=\Phi_{\circ}^2 c_{\text{L}} \xi /(16 \sqrt{2} \pi^2
\lambda^2 \gamma)$ and the collective pinning energy
$U_{\text{c}}=((\tilde{\gamma} \Phi_{\circ}^2 \xi^4)/(16 \pi^2
\lambda^2 \gamma^2 ))^{1/3}$. Here, $\Phi_{\circ}$ is the flux
quantum and $c_{\text{L}} \approx 0.1-0.2$ the Lindemann number.
$H_{\text{ss}}$ is inversely proportional to both the anisotropy
$\gamma$ and the disorder parameter $\tilde{\gamma}$. Concerning
the temperature dependence, as long as we are below the depinning
temperature $T_{\text{dp}} \approx U_{\text{c}}/k_{\text{B}}$, we
get \begin{equation} H_{\text{ss}} \propto \xi^{-3} \propto
(1-(T/T_{\text{c}})^4)^{3/2} \label{deltaTc} \end{equation} in the
case of $\delta T_{\text{c}} $-pinning and
\begin{equation}H_{\text{ss}} \propto \xi \propto
(1-(T/T_{\text{c}})^4)^{-1/2}\label{deltal} \end{equation} in the
case of $\delta \ell$-pinning.\cite{Twofluid} The calculation of
Ref.\ \onlinecite{Ertas96} leads to qualitatively the same
results. Since $H_{\text{ss}}$ decreases with temperature in our
case, the pinning in YBa$_{2-x}$Sr$_x$Cu$_4$O$_8$ cannot be
$\delta \ell$ pinning. The temperature dependence of
$H_{\text{ss}}$ of the unsubstituted crystal agrees satisfactorily
with the formula proposed for $\delta T_{\text{c}}$ pinning. In
the case of the substituted crystals, however, the agreement is
limited to values of $T/T_{\text{c}}$ between about $0.65$ and
$0.9$ (see Fig.\ \ref{Fig5}, the full bold lines are fits to Eq.\
(\ref{deltaTc})). The lowered $H_{\text{ss}}$ in the vicinity of
$T_{\text{c}}$ may be attributed to the finite transition width.

However, at low temperatures, the temperature dependence is rather
exponential in the case of the substituted crystals, as indicated
by dashed lines in Fig.\ \ref{Fig5}, with some indications of a
flattening at the lowest temperatures measured. This behaviour is
clearly at odds with Eq.\ (\ref{deltaTc}) and to the best
knowledge of the authors is not predicted by any present theory of
the order-disorder transition. An exponential upturn in the {\em
onset} field $H_{\text{on}}^{\uparrow}$ at low temperatures was
also found on a Nd$_{2-x}$Ce$_x$CuO$_{4-x}$ single crystal by
Andrade \textit{et al.},\cite{Andrade98} who attributed it to
Bean-Livingston surface barriers. However, since the observed
magnetic hysteresis is the sum of bulk and barrier hysteresis, it
is difficult to imagine why the kink in $M(H)$ at $H_{\text{ss}}$
should be influenced by surface or geometrical barriers, and in
our case a clear kink is observable at low temperatures where
$H_{\text{ss}}$ has already clearly departed from the
$(1-(T/T_{\text{c}})^4)^{3/2}$ dependence observed at higher
temperatures.

\begin{figure}
\includegraphics[width=0.95\linewidth]{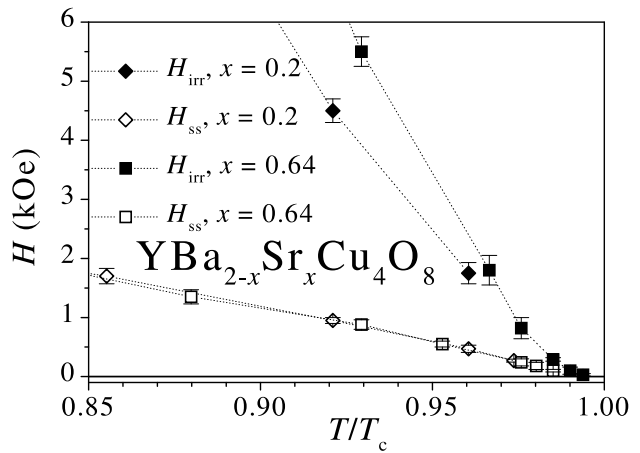}
\caption{$H_{\text{ss}}$ and $H_{\text{irr}}$ in the vicinity of
$T_{\text{c}}$, for substitutions of 10\% and 32\% Sr.}
\label{Fig9}
\end{figure}

A possible cause for a changed temperature dependence is a
dimensional crossover. Due to the layered structure of the cuprate
superconductors, vortex lines should be thought of being composed
of stacks of ``pancake'' vortices.\cite{Blatter94} Only below the
2D/3D crossover field $H_{\text{2D}}\approx \Phi_{\circ}/(\gamma^2
s^2)$, where $s$ is the distance between two adjacent layers, is
the interlayer interaction between pancake vortices larger than
their intralayer interaction and the pancakes form well defined
vortex lines. For all our crystals, $H_{\text{2D}}\gtrsim 77 \,
{\text{kOe}}$ is estimated well above the upper limit of the
fields attainable in our magnetometers. Also, $H_{\text{ss}}
\propto \xi^{-5/2}(\xi^{5/2})$ for 2D and $\delta
T_{\text{c}}(\delta \ell)$ pinning,\cite{Vinokur98} i.e. the
temperature dependence should be even flatter in the 2D regime. A
dimensional crossover can therefore not be responsible for the
observed upturn of $H_{\text{ss}}$ at low temperatures. In
principle a second pinning mechanism, which is very effective at
low temperatures, could be overlaid. However, point-like disorder
should rather (additionally) suppress $H_{\text{ss}}$, while the
influence of correlated disorder should be more visible at high
temperatures or low fields.\cite{Kwok00} Figure \ref{Fig9} shows
$H_{\text{ss}}$ and $H_{\text{irr}}$ of two of the substituted
crystals in the vicinity of the transition temperature. No sign of
a tricritical point, where the two lines would meet, can be seen.
A tricritical point would have to be located very near
$T_{\text{c}}$, where reliable measurements become increasingly
difficult.

It can be seen in Fig.\ \ref{Fig5} that any substitution of Sr
lowers $H_{\text{ss}}$ very significantly. For two values of
$T/T_{\text{c}}$, $H_{\text{ss}}$ vs.\ strontium content $x$ is
shown in Fig.\ \ref{Fig7}. After a substitution of just 10\% Sr
for Ba, $H_{\text{ss}}$ drops roughly by a factor of 5, at both
temperatures. A further increase of Sr substitution, however, does
not reduce the magnitude of $H_{\text{ss}}$ any more. Rather,
$H_{\text{ss}}(x)$ remains essentially constant. We do not
consider the peak at $x \approx 0.4$ as big enough to be
significant.

The initial decrease of $H_{\text{ss}}$ upon Sr substitution does
not come unexpected. It follows from Eq.\ (\ref{Hss_pos}) that
$H_{\text{ss}}\propto (\gamma \tilde{\gamma})^{-1}$. Sr
substitution increases the disorder parameter and decreases the
anisotropy (see Sec.\ \ref{Sr_disc}). The large decrease of
$H_{\text{ss}}$ upon substituting 10\% of Sr indicates that the
influence of the Sr substitution on disorder is much bigger than
it's influence on the anisotropy.

\begin{figure}[t!b]
\includegraphics[width=0.95\linewidth]{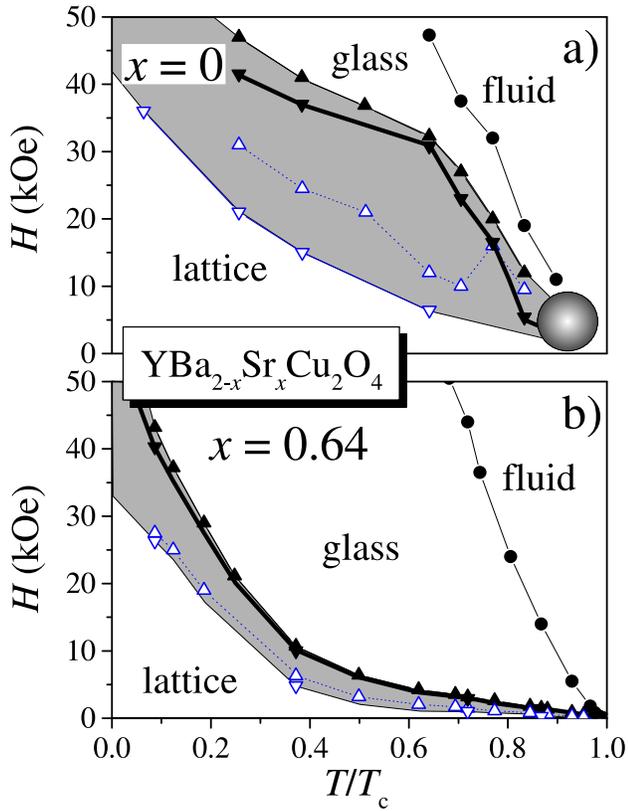}
\caption{Phase diagrams of the crystals with the lowest ($x=0$,
a)) and the highest ($x=0.64$, b)) Sr substitution. The
quasi-ordered lattice phase at low temperatures and fields is
separated from the highly disordered glass phase by the transition
line $H_{\text{ss}}^{\downarrow}$ (full down triangles, thick
line). Around the transition is a region of
metastability,\protect\cite{Radzyner00} where both phases can
coexist (shaded). The lower limit of metastability is marked by
$H_{\text{on}}^{\downarrow}$ (open down triangles) and the upper
limit, very close to the equilibrium transition, by
$H_{\text{ss}}^{\downarrow}$ (full up triangles). Also shown is
$H_{\text{on}}^{\uparrow}$ (open up triangles). The
irreversibility field $H_{\text{irr}}$ (full circles) is assumed
to correspond roughly to the transition line between glass and
fluid phases. Both $H_{\text{ss}}$ and $H_{\text{irr}}$ are
decreasing monotonically with increasing temperature. For the
crystal with $x=0$ the region of a possible tricritical point was
not measured and is blurred in the figure. For the crystal with
$x=0.64$ a tricritical point could not be detected and would have
to be located very near $T_{\text{c}}$.} \label{Fig10}
\end{figure}

What is more surprising is the apparent saturation of the
influence of the additional disorder for a substitution level $x
\gtrsim 0.2$, while the maximum pinning force is still linearly
increasing up to $x \approx 0.4$. In the case of weak point-like
pinning $H_{\text{ss}} \propto \tilde{\gamma}^{-1} \propto
n_{\text{dis}}^{-1}$, where $\tilde{\gamma}$ is the disorder
parameter and $n_{\text{dis}}$ the defect density, is
expected.\cite{Blatter94} It would seem natural to assume the
defect density to be proportional to the Sr substitution level
$x$. The blocking layer thickness on the other hand, which is the
structural factor with the biggest influence on the anisotropy,
decreases linearly with $x$.\cite{Karpinski00a} A large initial
decrease of $H_{\text{ss}}$ followed by an almost independence on
$x$ is therefore difficult to explain.

A downward shift of $H_{\text{ss}}$ upon introducing more disorder
was also found in electron irradiated crystals of
BSCCO\cite{Khaykovich97} and Y123.\cite{Nishizaki00} The notable
difference between both of these irradiation experiments and Sr
substitution in Y124 is that in both cases of electron irradiation
there is no sign of any saturation of the disorder induced
lowering of $H_{\text{ss}}$. As discussed above, the disorder
induced by Sr substitution is probably, even for the lowest
substitution level of $x \approx 0.2$, already much higher than
the disorder induced by the electron irradiation with the largest
fluence measured in Refs.\ \onlinecite{Khaykovich97} and
\onlinecite{Nishizaki00}. It may be that in our substituted
crystals the disorder is already too big for the applicability of
formulas derived for weak disorder.

The resulting phase diagrams for YBa$_{2}$Cu$_4$O$_8$ and
YBa$_{1.36}$Sr$_{0.64}$Cu$_4$O$_8$ are contrasted in Fig.\
\ref{Fig10}. The glass phase of YBa$_{1.36}$Sr$_{0.64}$Cu$_4$O$_8$
has, as compared to YBa$_2$Cu$_4$O$_8$, expanded dramatically from
a rather small part of the easily experimentally accessible phase
diagram to the phase covering the biggest area in fields $H \leq
55 \, {\text{kOe}}$. The most dramatic shift is the lowering of
$H_{\text{ss}}$, which is to be expected as the (quasi-)ordered
phase should become smaller upon introducing more disorder into
the system. The region of metastability is much smaller for the
substituted crystal, indicating that strong disorder tends to
reduce over-heating and especially under-cooling effects. We
speculate that this may be connected to our observation that for
unsubstituted Y124 with very weak disorder $H_{\text{ss}}$ depends
strongly on the exact amount of disorder, while for substituted
Y124 with rather strong disorder it almost does not depend on the
exact level of disorder. In real crystals, the disorder density
depends on the position on a mesoscopic scale. Soibel {\em et
al.}\cite{Soibel00} observed that for melting supercooling exists
only at local maxima of the transition field. In a first
approximation the width of metastability can be linked to the
difference between maximal and minimal transition fields within
the sample. In our case the same spatial variation of disorder
should produce a much bigger variation of the local transition
field for the unsubstituted crystal than for the substituted ones.

\section{Conclusions}
\label{conc}

We investigated the influence of structural disorder, introduced
by a substitution of Sr for Ba in YBa$_2$Cu$_4$O$_8$, on the
critical current density at elevated fields and on the borders of
the glass phase of vortices. We stress that the disorder
introduced chemically by means of Sr substitution is random and
point-like, rather than disorder due to extended defects. This is
shown by the hysteresis loops keeping their general form, similar
to the case of electron irradiation and unlike the case of neutron
irradiation, where at least partially the effect is to introduce
extended defects.

Our main conclusion is that the introduction of even a small
amount of disorder into very clean systems changes the phase
diagram drastically by lowering the order-disorder transition line
$H_{\text{ss}}(T)$ and also changing it's temperature dependence.
When the disorder reaches a certain threshold, however,
introducing additional disorder does not continue this tendency.
With other words, in highly disordered systems (including probably
most systems with partial chemical substitutions), the vortex
matter phase diagram is relatively robust with respect to
variations in the exact degree of disorder.

This conclusion is supported by the rather small variation of the
measured kink location as an indication of the order-disorder
transition at fixed reduced temperature for any Sr substitution
level $x \geq 0.2$, especially as compared to the corresponding
transition in a clean, unsubstituted crystal. That the disorder is
still increasing with increasing substitution level for $x>0.2$ is
indicated both by an increasing NQR line width\cite{Karpinski00a}
and increasing critical current densities at low fields and the
maximum pinning force density.

In strongly disordered systems, metastability is less important
than in very clean systems and especially the region of
undercooling of the glass phase is much smaller. This might be
linked to the observation that the position of the transition line
does not depend much on the exact amount of disorder in strongly
disordered systems, since variations of the disorder density on a
mesoscopic scale do not produce large positional differences of
the local transition field in this case.

Similar to the robustness of the phase diagram above a disorder
threshold, both the maximum pinning force density and the maximum
critical current density in the glass phase increase initially
with increasing disorder, the critical current density increases
by an order of magnitude upon substituting 10\% Sr. But they then
show saturation behaviour, i.e.\ the pinning force density and
critical current densities can be raised only up to a certain
point by introducing random point-like disorder. This may be
relevant for possible applications where high critical current
densities are required.

It should be noted that current detailed theories of the
order-disorder transition generally assume the disorder to be very
weak. Our measurements indicate that in systems with a high amount
of disorder, the phase diagram and in particular the
order-disorder transition differ qualitatively from the weak
disorder case. Theoretical investigations beyond the weak disorder
limit would be very helpful.

Both the pinning of unsubstituted Y124 and Sr substituted Y124 is
likely to be of $\delta T_{\text{c}}$, rather than $\delta \ell$
type, indicated by the temperature dependence of the
order-disorder transition at intermediate temperatures. For the
substituted crystals this can be expected, since there are
variations of $T_{\text{c}}$ upon Sr substitution. However, the
exponential-like temperature dependence of the order-disorder
transition of substituted crystals at low temperatures is also
emphasized. The dependence cannot be linked to the influence of
surface barriers, at least not in any straightforward way, the
same dependence of all significant fields (onsets and clearly
discernible kinks for both field directions) suggests it is a true
bulk property of the transition. We stress that the observed
temperature dependence is at odds with currently published
theoretical formulas.

\begin{acknowledgments}
\label{ack} This work was supported by the Swiss National Science
Foundation, by the European Community (program ICA1-CT-2000-70018,
Centre of Excellence CELDIS) and by the Polish State Committee for
Scientific Research (KBN, contract No. 5 P03B 12421).
\end{acknowledgments}









\newcommand{\noopsort}[1]{} \newcommand{\printfirst}[2]{#1}
  \newcommand{\singleletter}[1]{#1} \newcommand{\switchargs}[2]{#2#1}


\begin{thebibliography}{10}


\bibitem{Blatter94}
G. Blatter, M.~V. Feigel'man, V.~B. Geshkenbein, A.~I. Larkin, and
V.~M.
  Vinokur, Rev. Mod. Phys. {\bf 66},  1125  (1994).

\bibitem{Sasagawa98}
T. Sasagawa, K. Kishio, Y. Togawa, J. Shimoyama, and K. Kitazawa,
Phys. Rev.
  Lett. {\bf 80},  4297  (1998).

\bibitem{Safar92}
H. Safar, P.~L. Gammel, D.~A. Huse, D.~J. Bishop, J.~P. Rice, and
D.~M. Ginsberg, Phys. Rev. Lett. {\bf 69},  824  (1992); H. Safar,
P.~L. Gammel, D.~A. Huse, D.~J. Bishop, W.~C. Lee, J.
Giapintzakis, and D.~M. Ginsberg, {\em ibid}. {\bf 70},  3800
(1993).

\bibitem{Cubitt93}
R. Cubitt, E.~M. Forgan, G. Yang, S.~L. Lee, D.~M. Paul, H.~A.
Mook, M.
  Yethiraj, P.~H. Kes, T.~W. Li, A.~A. Menovsky, Z. Tarnawski, and K.
  Mortensen, Nature {\bf 365},  407  (1993).

\bibitem{Schilling96}
A. Schilling, R.~A. Fisher, N.~E. Phillips, U. Welp, D. Dasgupta,
W.~K. Kwok,
  and G.~W. Crabtree, Nature {\bf 382},  791  (1996).

\bibitem{Khaykovich96}
B. Khaykovich, E. Zeldov, D. Majer, T.~W. Li, P.~H. Kes, and M.
Konczykowski,
  Phys. Rev. Lett. {\bf 76},  2555  (1996).

\bibitem{Khaykovich97}
B. Khaykovich, M. Konczykowski, E. Zeldov, R.~A. Doyle, D. Majer,
P.~H. Kes,
  and T.~W. Li, Phys. Rev. B {\bf 56},  R517  (1997).

\bibitem{Deligiannis97}
K. Deligiannis, P.~A.~J. de~Groot, M. Oussena, S. Pinfold, R.
Langan, R.
  Gagnon, and L. Taillefer, Phys. Rev. Lett. {\bf 79},  2121  (1997).

\bibitem{Giller97}
D. Giller, A. Shaulov, R. Prozorov, Y. Abulafia, Y. Wolfus, L.
Burlachkov, Y.
  Yeshurun, E. Zeldov, V.~M. Vinokur, J.~L. Peng, and R.~L. Greene, Phys. Rev.
  Lett. {\bf 79},  2542  (1997).

\bibitem{Nishizaki98}
T. Nishizaki, T. Naito, and N. Kobayashi, Phys. Rev. B {\bf 58},
11169
  (1998).

\bibitem{Giller99}
D. Giller, A. Shaulov, Y. Yeshurun, and J. Giapintzakis, Phys.
Rev. B {\bf 60},
   106  (1999).

\bibitem{Nishizaki00}
T. Nishizaki, T. Naito, S. Okayasu, A. Iwase, and N. Kobayashi,
Phys. Rev. B
  {\bf 61},  3649  (2000).

\bibitem{Radzyner00}
Y. Radzyner, S.~B. Roy, D. Giller, Y. Wolfus, A. Shaulov, P.
Chaddah, and Y.
  Yeshurun, Phys. Rev. B {\bf 61},  14362  (2000).

\bibitem{Nattermann90}
T. Nattermann, Phys. Rev. Lett. {\bf 64},  2454  (1990); T.
Nattermann and S. Scheidl, Adv. Phys. {\bf 49},  607  (2000).

\bibitem{Giamarchi94}
T. Giamarchi and P. Le~Doussal, Phys. Rev. Lett. {\bf 72},  1530
(1994); Phys. Rev. B {\bf 52},  1242 (1995).

\bibitem{Fisher89}
M.~P.~A. Fisher, Phys. Rev. Lett. {\bf 62},  1415  (1989); D.~S.
Fisher, M.~P.~A. Fisher, and D.~A. Huse, Phys. Rev. B {\bf 43},
130  (1991).

\bibitem{Reichhardt00}
C. Reichhardt, A. van~Otterlo, and G.~T. Zim{\'a}nyi, Phys. Rev.
Lett. {\bf 84},
   1994  (2000).

\bibitem{Nelson92}
D.~R. Nelson and V.~M. Vinokur, Phys. Rev. Lett. {\bf 68},  2398
(1992); Phys. Rev. B {\bf 48},  13060  (1993).

\bibitem{Kwok00}
W.~K. Kwok, R.~J. Olsson, G. Karapetrov, L.~M. Paulius, W.~G.
Moulton, D.~J.
  Hofman, and G.~W. Crabtree, Phys. Rev. Lett. {\bf 84},  3706  (2000).

\bibitem{Vinokur98}
V. Vinokur, B. Khaykovich, E. Zeldov, M. Konczykowski, R.~A.
Doyle, and P.~H.
  Kes, Physica C {\bf 295},  209  (1998).

\bibitem{Ertas96}
D. Ertas and D.~R. Nelson, Physica C {\bf 272},  79  (1996).

\bibitem{Giamarchi97}
T. Giamarchi and P. Le~Doussal, Phys. Rev. B {\bf 55},  6577
(1997).

\bibitem{Koshelev98}
A.~E. Koshelev and V.~M. Vinokur, Phys. Rev. B {\bf 57},  8026
(1998).

\bibitem{Gaifullin00}
M.~B. Gaifullin, Y. Matsuda, N. Chikumoto, J. Shimoyama, and K.
Kishio, Phys.
  Rev. Lett. {\bf 84},  2945  (2000).

\bibitem{Giller00}
D. Giller, A. Shaulov, T. Tamegai, and Y. Yeshurun, Phys. Rev.
Lett. {\bf 84},
  3698  (2000).

\bibitem{Paltiel00}
Y. Paltiel, E. Zeldov, Y.~N. Myasoedov, H. Shtrikman, S.
Bhattacharya, M.~J. Higgins, Z.~L. Xiao, E.~Y. Andrei, P.~L.
Gammel, and D.~J. Bishop, Nature {\bf 403},  398  (2000); Y.
Paltiel, E. Zeldov, Y. Myasoedov, M.~L. Rappaport, G. Jung, S.
Bhattacharya, M.~J. Higgins, Z.~L. Xiao, E.~Y. Andrei, P.~L.
Gammel, and D.~J. Bishop, Phys. Rev. Lett. {\bf 85},  3712
(2000).

\bibitem{Avraham01}
N. Avraham, B. Khaykovich, Y. Myasoedov, M. Rappaport, H.
Shtrikman, D.~E.
  Feldman, T. Tamegai, P.~H. Kes, M. Li, M. Konczykowski, K. {van der Beek},
  and E. Zeldov, Nature {\bf 411},  451  (2001).

\bibitem{Karpinski99}
J. Karpinski, G.~I. Meijer, H. Schwer, R. Molinski, E. Kopnin, K.
Conder, M.
  Angst, J. Jun, S. Kazakov, A. Wisniewski, R. Puzniak, J. Hofer, V. Alyoshin,
  and A. Sin, Supercond. Sci. Technol. {\bf 12},  R1  (1999).

\bibitem{Dover89}
See e.g.~R.~B. {van}$\;$Dover, E.~M. Gyorgy, L.~F. Schneemeyer,
J.~W. Mitchell, K.~V. Rao, R. Puzniak, and J.~V. Waszczak, Nature
{\bf 342},  55  (1989); L. Civale, A.~D. Marwick, M.~W. McElfresh,
T.~K. Worthington, A.~P. Malozemoff, F.~H. Holtzberg, J.~R.
Thompson, and M.~A. Kirk, Phys. Rev. Lett. {\bf 65},  1164 (1990).

\bibitem{Murakami94}
See e.g.~M. Murakami, S.-I. Yoo, T. Higuchi, N. Sakai, J. Weltz,
N. Koshizuka, and S. Tanaka, Jpn. J. Appl. Phys. {\bf 33},  L715
(1994).

\bibitem{Karpinski00a}
J. Karpinski, S. Kazakov, M. Angst, A. Mironov, M. Mali, and J.
Roos, Phys.
  Rev. B {\bf 64},  094518 (2001).

\bibitem{notejcjs}
Actually rather the shielding current density $j_{\text{s}}(t=5 \,
  {\text{min}})$ than the true critical current density $j_{\text{c}} \equiv
  j_{\text{s}}(t=0)$, but since the time scale used is a typical one for SQUID
  measurements we still denote it $j_{\text{c}}$.

\bibitem{Wiesinger92}
H.~P. Wiesinger, F.~M. Sauerzopf, and H.~W. Weber, Physica C {\bf
203},  121
  (1992).

\bibitem{Attfield98}
J. Attfield, A. Kharlanov, and J. McAllister, Nature {\bf 394},
157  (1998).

\bibitem{ownfutur}
M. Angst {\textit{et al.}} (unpublished).

\bibitem{Marchetti91}
M.~C. Marchetti and D.~R. Nelson, Physica C {\bf 174},  40
(1991).

\bibitem{Abulafia96}
Y. Abulafia, A. Shaulov, Y. Wolfus, R. Prozorov, L. Burlachkov, Y.
Yeshurun, D.
  Majer, E. Zeldov, H. W{\"u}hl, V.~B. Geshkenbein, and V.~M. Vinokur, Phys.
  Rev. Lett. {\bf 77},  1596  (1996).

\bibitem{note2pbelowHm}
The observation of a prominent second peak at high temperatures
clearly in the lattice phase can be explained by thermal
production of a finite dislocation density. See Refs.\
\protect\onlinecite{Deligiannis97,Giamarchi97}.

\bibitem{Werner00}
M. Werner, F.~M. Sauerzopf, H.~W. Weber, and A. Wisniewski, Phys.
Rev. B {\bf
  61},  14795  (2000).

\bibitem{Werner98}
M. Werner, G. Brandst{\"a}tter, F.~M. Sauerzopf, H.~W. Weber, A.
Hoekstra, R.
  Surdeanu, R.~J. Wijngaarden, R. Griessen, Y. Abulafia, Y. Yeshurun, K.
  Winzer, and B.~W. Veal, Physica C {\bf 303},  191  (1998).

\bibitem{Bean_Livingston64}
C.~P. Bean and J.~D. Livingston, Phys. Rev. Lett. {\bf 12},  14
(1964).

\bibitem{Burlachkov94}
L. Burlachkov, V.~B. Geshkenbein, A.~E. Koshelev, A.~I. Larkin,
and V.~M.
  Vinokur, Phys. Rev. B {\bf 50},  16770  (1994).

\bibitem{Zeldov94}
E. Zeldov, A.~I. Larkin, V.~B. Geshkenbein, M. Konczykowski, D.
Majer, B. Khaykovich, V.~M. Vinokur, and H. Shtrikman, Phys. Rev.
Lett. {\bf 73},  1428  (1994); E. Zeldov, A.~I. Larkin, M.
Konczykowski, B. Khaykovich, D. Majer, V.~B. Geshkenbein, and
V.~M. Vinokur, Physica C {\bf 235-240},  2761  (1994).

\bibitem{Doyle98}
R.~A. Doyle, S.~F. W.~R. Rycroft, C.~D. Dewhurst, E. Zeldov, I.
Tsabba, S.
  Reich, T.~B. Doyle, T. Tamegai, and S. Ooi, Physica C {\bf 308},  123
  (1998).

\bibitem{Zeldov95b_Dewhurst96}
E. Zeldov, D. Majer, M. Konczykowski, A.~I. Larkin, V.~M. Vinokur,
V.~B. Geshkenbein, N. Chikumoto, and H. Shtrikman, Europhys. Lett.
{\bf 30},  367  (1995); C.~D. Dewhurst, D.~A. Cardwell, A.~M.
Campbell, R.~A. Doyle, G. Balakrishnan, and D.~M. Paul, Phys. Rev.
B {\bf 53},  14594  (1996).

\bibitem{Yamauchi98}
See e.g.~H. Yamauchi, M. Karppinen, K. Fujinami, T. Ito, H.
Suematsu, K. Matsuura, and
  K. Isawa, Supercond. Sci. Technol. {\bf 11},  1006  (1998).

\bibitem{Werner97}
M. Werner, Ph.D. thesis, Technical University Vienna, 1997.

\bibitem{Ikeda96}
R. Ikeda, J. Phys. Soc. Jpn. {\bf 65},  3998  (1996).

\bibitem{Larkin72}
A.~I. Larkin and Y.~N. Ovchinnikov, Sov. Phys. JETP {\bf 34},  651
(1972); J. Low Temp. Phys. {\bf 34}, 409  (1979).

\bibitem{Twofluid}
We used the two-fluid approximation for the temperature dependence
of the
  superconducting parameters.

\bibitem{Andrade98}
M.~C. de~Andrade, N.~R. Dilley, F. Ruess, and M.~B. Maple, Phys.
Rev. B {\bf
  57},  R708  (1998).

\bibitem{Soibel00}
A. Soibel, E. Zeldov, M. Rappaport, Y. Myasoedov, T. Tamegai, S.
Ooi, M. Konczykowski, and V.~B. Geshkenbein, Nature {\bf
  406},  282  (2000).

\end{thebibliography}
\end{document}